\documentclass[11pt, twoside, a4paper]{article}
\usepackage[pdftex]{graphicx}
\pdfoutput=1
\usepackage{tabularx}
\usepackage{amsmath}\usepackage{amsfonts}
	\numberwithin{equation}{section}
    
\newcommand{\norm}[1]{\left\lVert#1\right\rVert}
\usepackage{chngcntr}
\counterwithin{figure}{section}
\counterwithin{table}{section}

\usepackage[super]{nth}

\usepackage{booktabs}
\usepackage{algorithm}
\usepackage{multirow}
\usepackage{xfrac}
\usepackage{algpseudocode}
\usepackage{placeins}
\usepackage{layouts}
\usepackage{pdflscape}
\usepackage{blindtext}
\usepackage{bm}
\usepackage{tikz}
\usepackage[round]{natbib}
\usepackage{textcomp}
\usepackage[english]{babel}
\usepackage{latexsym}
\usepackage{mathtools}
\usepackage{caption}
\usepackage{pdfpages}
\usepackage{listings}
\usepackage{hyperref}
\usepackage{flafter}

\makeatletter
\newcommand{\algorithmfootnote}[2][\footnotesize]{%
  \let\old@algocf@finish\@algocf@finish
  \def\@algocf@finish{\old@algocf@finish
    \leavevmode\rlap{\begin{minipage}{\linewidth}
    #1#2
    \end{minipage}}%
  }%
}
\makeatother
    
\usepackage{siunitx}
\usepackage{resizegather}
\usepackage{epstopdf}
\usepackage{color}
\definecolor{mygray}{rgb}{0.5, 0.5, 0.5}
\let\orgautoref\autoref
\providecommand{\Autoref}
        {\def\equationautorefname{Equation}%
         \def\figureautorefname{Figure}%
         \def\subfigureautorefname{Figure}%
         \def\Itemautorefname{Item}%
         \def\tableautorefname{Table}%
         \def\sectionautorefname{Section}%
         \def\subsectionautorefname{Section}%
         \def\subsubsectionautorefname{Section}%
         \def\chapterautorefname{Section}%
         \def\partautorefname{Part}%
         \orgautoref}

\renewcommand{\Re}{\operatorname{Re}}
\renewcommand{\Im}{\operatorname{Im}}

\makeatletter
\renewcommand*\env@matrix[1][\arraystretch]{%
  \edef\arraystretch{#1}%
  \hskip -\arraycolsep
  \let\@ifnextchar\new@ifnextchar
  \array{*\c@MaxMatrixCols c}}
\makeatother

\newcommand{\overbar}[1]{\mkern 1.5mu\overline{\mkern-1.5mu#1\mkern-1.5mu}\mkern 1.5mu}

\newcommand{\absm}{\lvert m \rvert}

\begin{document}

\setcounter{page}{1}
\begin{center}
{\bf  Implementation and benchmarking of a crosstalk-free method for wavefront Zernike coefficients reconstruction using Shack-Hartmann sensor data}\\
R. S. Biesheuvel$^1$, A. J. E. M. Janssen$^2$, P. Pozzi$^3$, and S. F. Pereira$^1$\\
{\it $^1$ Optics Research Group, ImPhys Department\\
Faculty of Applied Sciences, Delft University of Technology\\
Lorentzweg 1, 2628 CJ Delft, The Netherlands\\
$^2$ Department of Mathematics and Computer Science\\
 Eindhoven University of Technology\\
P.O.Box 513, 5600 MB Eindhoven, The Netherlands\\
$^3$ Delft Center of Systems and Control\\
Delft University of Technology\\
Mekelweg 2, 2628 CD Delft, The Netherlands}

email: s.f.pereira@tudelft.nl\\
\end{center} 
\begin{abstract}
In wavefront characterization, often the combination of a Shack-Hartmann sensor and a reconstruction method utilizing the Cartesian derivatives of Zernike circle polynomials (the least-squares method, to be called here Method A) is used, which is known to introduce crosstalk. In \citep{janssen2014zernike} a crosstalk-free analytic expression of the LMS estimator of the wavefront Zernike coefficients is given in terms of wavefront partial derivatives (leading to what we call Method B). Here, we show an implementation of this analytic result where the derivative data are obtained using the Shack-Hartmann sensor and compare it with the conventional least-squares method.

\end{abstract}
\setcounter{tocdepth}{3}
\pagenumbering{arabic}
\section{Introduction}

In applications involving wavefront reconstruction, Shack-Hartmann sensors are frequently used. An often-used polynomial basis to expand the phase of the scalar field in an optical system is the set of Zernike circle polynomials. The wavefront Zernike coefficients can be estimated with for example, the least-squares method, to be called Method A henceforth. However, already since the 1980s, it is known that there is a fundamental problem in the combination of the least-squares method and Zernike polynomials: this method presents crosstalk between coefficients. In 2014, Janssen introduced an analytic result in the form of a new relation between the wavefront derivative data and the wavefront Zernike coefficients \citep{janssen2014zernike}. This relation theoretically solves the crosstalk problem that is present in the least-squares method. In particular, the wavefront coefficients, once estimated, do not change anymore when the number of Zernike circle polynomials involved in the fit is increased further.

In this paper, we show an implementation of Janssen's analytic result in the reconstruction of several wavefronts that have been generated using a spatial light modulator. In particular, we show that this method, to be called Method B henceforth, does not suffer from cross-talk between the coefficients.

The paper is organized as follows. In \Autoref{sec:theory} we present theoretical aspects of the problem, and in \autoref{sec:setup2_method2}, the experimental setup and methods are explained. The actual results are presented and discussed in \autoref{sec:results}, and the conclusions from these measurements are presented in \autoref{sec:conclusion}. 
\section{Theory} \label{sec:theory}
In this section, Zernike circle polynomials are discussed and used as a basis to describe the  wavefront deviation with respect to a suitably defined reference sphere of an optical system. This wavefront deviation or wavefront aberration function will be loosely called 'wavefront aberration' or 'aberration' in what follows. We start this section by showing the definition of Zernike polynomials and a description of the data that is obtained with the Shack-Hartmann sensor is discussed. Next, we discuss two reconstruction methods: the well-known least-squares method (Method A) and the recently method introduced by Janssen\citep{janssen2014zernike} (Method B). 

\subsection{Zernike polynomials as a basis for wavefront expansion}
In systems where aberrations are desired to be known, such as microscopes and telescopes, a circular aperture is usually present. In order to describe these aberrations, Zernike circle polynomials are commonly used. In this paper, the American National Standards Institute (ANSI) definition for Zernike circle polynomials is considered, which is commonly used to describe wavefront reconstruction from Shack-Hartmann data.\footnote{The choice for ANSI convention is motivated here by how the Shack-Hartmann sensor deals with the circle polynomials. We would like to point out that using the Born and Wolf convention (that uses exponential instead of cosine/sine functions), renders the results in \citep{janssen2014zernike} more transparent and concise.} As defined in \citet{thibos2002standards}, the polynomials in polar coordinates $(\rho, \theta)$ are given by

\begin{equation}
\label{eq:radial_fact}
z_n^m(\rho, \theta) = N_n^m R_n^{\lvert m \rvert}(\rho) \Theta_m(\theta),
\end{equation}
where
\begin{align}
N_n^m &= \sqrt{(2 - \delta_{m0}) (n+1)} \label{eq:nnm},\\
R_n^{\lvert m \rvert}(\rho) &= \sum_{s=0}^{\frac{n - \lvert m \rvert}{2}} \frac{(-1)^s (n-s)!}{s! \left( \frac{n - \lvert m \rvert}{2} - s \right)! \left( \frac{n + \lvert m \rvert}{2} -s \right)!} \rho^{n - 2s},\\
\Theta_m(\theta) &= \begin{cases}
    \cos(m \theta),& \text{if } m\geq 0,\\
    -\sin(m \theta),& \text{if } m<0,
\end{cases}
\end{align}
such that $z_n^m (\rho, \theta)$ is a real-valued, orthonormal (on the unit disc) expression for Zernike circle polynomials, $n= 0,1,2,...$ is the degree of the Zernike polynomial, and $m=0,\pm 1,\pm 2,...$ the azimuthal order satisfying:
\begin{align}\label{eq:nbigger0}
&n \geq 0,\\
&n - \lvert m \rvert~ \text{is~even}, \label{eq:evennm}\\ \label{eq:msmallern}
&\lvert m \rvert \leq n, 
\end{align}
$\rho \leq 1$ being the radius on the unit disc, and $\delta_{nn'}$ the Kronecker delta function. Here, the $n^{\text{th}}$ ``order'' polynomial is referred to as the $n^{\text{th}}$ \emph{degree} polynomial. 

As mentioned above, these circle polynomials are orthonormal on the unit disc, i.e.,

\begin{equation}
\frac{1}{\pi}\int_0^1 \int_0^{2 \pi} z_n^m(\rho, \theta) z_{n'}^{m'}(\rho, \theta) \rho \mathrm{d}\theta \mathrm{d}\rho = \delta_{nn'} \delta_{mm'}. 
\end{equation}

With this definition, any real-valued wavefront function (defined on the unit disc) can be described as a linear combination of the Zernike circle polynomials given by
\begin{equation}
\label{eq:wavefront_sum}
W(\rho, \theta) = \sum_{m=-\infty}^{\infty} \sum_{n \in \eta_m} a_{n}^m z_n^m(\rho, \theta),
\end{equation}
where $\eta_m$ is the set of allowed $n$ values dependent on $m$, namely $\eta_m = \left\{ \absm, \absm + 2, \absm + 4, \ldots \right\}$, when $m \neq 0$, and $\eta_m = \left\{ 2, 4, \ldots, \infty \right\}$ when $m = 0$ and $z_0^0=1$. This set ensures that the constraints set in $n$ and $m$ are all met. Finally, $a_n^m$ is the real-valued Zernike coefficient.

\subsection{Complex-valued definition of the Zernike circle polynomials}
Following the complex-valued definition of the Zernike circle polynomials as used in \citep{janssen2014zernike} in polar coordinates $(\rho, \theta)$ on the unit disc, we have that
\begin{equation}
\label{eq:complex_def}
Z_n^m(\rho, \theta) = R_n^{\lvert m \rvert}(\rho)e^{im\theta}.
\end{equation}
The radial polynomial $R_n^{\lvert m \rvert}$ is given as in \Autoref{eq:radial_fact} above, and can also be given as
\begin{equation}
\label{eq:rmn}
R_n^{\lvert m \rvert}(\rho) = \rho^{\absm}P^{(0,\lvert m \rvert)}_{\frac{n - \lvert m \rvert}{2}}(2\rho^2 - 1),
\end{equation}
where $P_k^{(\alpha, \beta)}(x)$ is the Jacobi polynomial of degree $k$, which is orthogonal with respect to the weight $(1-x)^\alpha (1+x)^\beta$ on the interval $[-1, 1]$. Note that the factor $\rho^{\absm}$ is missing in \citep{janssen2014zernike}, Eq. (6), but it is merely a typo and has no consequences in the further developments in that reference. We set $Z_n^m = 0$ for all values of $n$ and $m$ where $n - \lvert m \rvert$ is odd or negative. There is the normalization condition 
\begin{equation}
\label{eq:WW}
\int_0^1 \int_0^{2 \pi} Z_n^m(\rho, \theta) \left(Z_{n'}^{m'}(\rho, \theta)\right)^* \rho \mathrm{d}\theta \mathrm{d}\rho = \frac{\pi}{n+1}\delta_{nn'} \delta_{mm'}.
\end{equation}
This orthogonality means that, like the real-valued Zernike polynomial, any sufficiently smooth (complex) wavefront can be described by
\begin{equation} 
\label{eq:alpha}
W(\rho, \theta) =  \sum_{m=-\infty}^{\infty} \sum_{n \in \eta_m} \alpha_{n}^m Z_n^m(\rho, \theta),
\end{equation}
where $\alpha_n^m$ are generally complex-valued coefficients corresponding to the complex Zernike polynomial and $\eta_n$ is as in \Autoref{eq:wavefront_sum}.
Expanding the complex exponential in \Autoref{eq:complex_def} leads to the following conversion between this definition of complex Zernike polynomials and the ANSI standard
\begin{equation}
N_n^m Z_n^m(\rho, \theta) = \begin{dcases}z_n^{\absm}(\rho, \theta) + i z_n^{-\absm}(\rho, \theta),& \text{if}~m>0,\\
z_n^{\absm}(\rho, \theta) - i z_n^{-\absm}(\rho, \theta),& \text{if}~m<0,\\
z_n^m, & \text{if}~m=0,
\end{dcases}
\end{equation}
where $N_n^m$ is defined in \Autoref{eq:nnm}.

From the definition of the complex Zernike polynomial in \Autoref{eq:complex_def} we can also see that the complex conjugate $\left({Z_n^{\absm}}\right)^{*}$ is equal to $Z_n^{-\absm}$. This observation, together with the relations between the real and complex Zernike polynomials leads to the expression
\begin{align}
\label{eq:z_to_c}
z_n^m &=
\begin{dcases}
\frac{N_n^{m}}{2}\left(Z_n^{\absm} + Z_n^{-\absm}\right) = N_n^m \text{Re}\left(Z_n^{\absm}\right), & \text{if}~m > 0\\
\frac{N_n^{m}}{2i}\left(Z_n^{\absm} - Z_n^{-\absm}\right) = N_n^m \text{Im}\left(Z_n^{\absm}\right), & \text{if}~m < 0\\
N_n^{m}Z_n^{\absm}, & \text{if}~m=0.
\end{dcases}
\end{align}

When using complex polynomials to describe a real wavefront, the coefficients are usually complex valued. In order to transform them back to the coefficients of the real-valued Zernike $z_n^m$, the following relations can be used
\begin{align}
\label{eq:alphatoa}
a_n^m &= \begin{dcases} \frac{1}{N_n^m} \Re \left( \alpha_n^{\absm} + \alpha_n^{-\absm}\right), & \text{if}~m>0\\
\frac{-1}{N_n^m} \Im \left( \alpha_n^{\absm} - \alpha_n^{-\absm}\right), & \text{if}~m<0\\
\frac{1}{N_n^m} \Re \left( \alpha_n^m \right),& \text{if}~m=0.
\end{dcases} 
\end{align}
Here, the Zernike polynomials are evaluated on the computer, and data is saved in vectors and matrices. To make it easy to loop over all polynomials, a single index is introduced. In the notation by \citet{thibos2002standards}, this single index $j= 0,1,...$ is given by
\begin{equation}
\label{eq:j}
j = \frac{n(n+2) + m}{2},
\end{equation}
where $n$ is the Zernike degree and $m$ the azimuthal order. From now on, we will write $Z_j=z_n^m$. The reverse can also be done, that is finding $n$ and $m$ from $j$ as follows
\begin{align}
\label{eq:n}
n &= \left\lceil \frac{-3 + \sqrt{9+8j}}{2} \right\rceil, \\
\label{eq:m}
m &= 2j - n(n+2),
\end{align}
where $\left\lceil x \right\rceil$ denotes the ``ceiling'' function, that is the smallest integer greater or equal than $x$.

\subsection{Data from the Shack-Hartmann sensor} \label{sec:sh_sensor}
It is well known that aberrations in an optical system degrade its imaging quality. The phase aberration is defined by a function $\varphi_W = arg(W)$. A typical way of measuring this function is by using a Shack-Hartmann sensor. A Shack-Hartmann sensor consists of a camera chip and lenslet array. The lenslets are placed at the focal distance from the camera chip such that an incoming plane wave will be focussed as many spots on the camera. If the incoming wave is aberrated, the position of each spot on the sensor will changed.

It can be derived from the theory how much a spot is displaced due to an aberration. Because this displacement is linear, the reverse problem can be solved as well. In other words, one can retrieve the aberration given the spot displacement.

\citet{dai2008wavefront} has proven that for a varying wavefront over the subaperture, the slope needs to be averaged over that sub-aperture. The expressions for the average slopes in terms of displacement of the spot on the camera becomes
\begin{equation} \label{eq:disp_W}
\begin{dcases}
\frac{1}{A_{\Sigma}}{}\int_{\Sigma}\frac{\partial W}{\partial x}\mathrm{d}x\mathrm{d}y &= r \frac{\Delta x}{f}\\
\frac{1}{ A_{\Sigma}} \int_{\Sigma} \frac{\partial W}{\partial y}\mathrm{d}x \mathrm{d}y &= r \frac{\Delta y}{f} \\
\end{dcases}
\end{equation}
where $\Delta x, \Delta y$ are the shift in the $x-$ and $y-$ positions of the spots, $r$ is the radius of the incoming beam on the Shack-Hartmann sensor, and $f$ is the focal length of the sensor lens array, $\Sigma$ is the illuminated sub-aperture domain, with surface area $A_{\Sigma}$. Here $\Sigma$ and $A_{\Sigma}$ change when the sub-aperture is only partially illuminated (i.e., at the edge of the beam). This averaging of the slope is important in recovering the wavefront.

\subsubsection{Method A} \label{sec:lsq_theory}
The least-squares (LSQ) fit is based on the real Zernike polynomials, and uses the fact that the coefficients are not dependent on $x$ and $y$. The wavefront in $x, y$ coordinates (with $\sqrt{x^2 + y^2} \leq 1$) can be described as

\begin{equation}
W(x, y) = \sum_{m = -\infty}^{\infty} \sum_{n \in \eta_m} a_n^m z_n^m(x, y).
\end{equation}

The measured quantity, however, is not $W$ but $\frac{\partial W}{\partial x}$ and $\frac{\partial W}{\partial y}$. Taking the partial derivatives to $x$ and $y$ results in the over determined system of

\begin{equation}\label{eq:unsolved_system}
\begin{dcases}
\frac{\partial W}{\partial x} = \sum_{m = -\infty}^{\infty} \sum_{n \in \eta_m} a_n^m \frac{\partial z_n^m}{\partial x}\\
\frac{\partial W}{\partial y} = \sum_{m = -\infty}^{\infty} \sum_{n \in \eta_m} a_n^m \frac{\partial z_n^m}{\partial y}.
\end{dcases}
\end{equation}

This system can be solved for a finite set of polynomials. Using \autoref{eq:disp_W}, given the displacements $\Delta x$ and $\Delta y$, one can create a vector $\mathbf{s}$ containing the slopes as such
\begin{equation} \label{eq:slopes_vec}
\mathbf{s} = \begin{bmatrix}
\left. \overbar{\dfrac{\partial W}{\partial x}} \right \vert_{1} & %
\left. \overbar{\dfrac{\partial W}{\partial x}} \right \vert_{2} & %
\cdots & %
\left. \overbar{\dfrac{\partial W}{\partial x}} \right \vert_{n_{\text{spots}}} & %
\left. \overbar{\dfrac{\partial W}{\partial y}} \right \vert_{1} & %
\left. \overbar{\dfrac{\partial W}{\partial y}} \right \vert_{2} &%
\cdots & \left. \overbar{\dfrac{\partial W}{\partial y}} \right \vert_{n_{\text{spots}}}\\
\end{bmatrix}^{T}.
\end{equation}
The partial derivatives of the Zernikes in the $x$- and $y$-direction can also be put in a matrix, called the geometry matrix. We recall the convention that $Z_j=z_n^m$ with $j,m,n$ related as in \Autoref{eq:j} - \Autoref{eq:m}. The geometry matrix $G$ can be built up $\left. \overbar{ \dfrac{\partial Z_j}{\partial x}} \right \vert_{n}$, the average gradient of Zernike mode $j$ at the position of subaperture $n$. This averaging is done due to the fact that the spot displacement measured with the Shack-Hartmann sensor is proportional to the average slope of the wavefront, as expressed in \autoref{eq:disp_W}. It should be noted that the positions over which the averaging is done is normalized to the unit disc. These windows are the same defined in \autoref{eq:disp_W}. The matrix will have a size of $(2n_{\text{spot}} \times J)$, where $J$ is the maximum index for the Zernike modes used to have a good approximation of the true wavefront. The expression for $G$ becomes
\begin{equation} \label{eq:geo_mat}
G = \begin{bmatrix}[1.5]
\left. \overbar{ \dfrac{\partial Z_1}{\partial x}} \right \vert_{1}& \left.\overbar{ \dfrac{\partial Z_1}{\partial x}}\right\vert _{2} & \cdots & \left.\overbar{ \dfrac{\partial Z_1}{\partial x}}\right\vert _{n_{\text{spot}}} & 
\left.\overbar{ \dfrac{\partial Z_1}{\partial y}}\right\vert _{1} &
\left.\overbar{ \dfrac{\partial Z_1}{\partial y}}\right\vert _{2} &
\cdots & \left.\overbar{ \dfrac{\partial Z_1}{\partial y}}\right\vert _{n_{\text{spot}}} \\
\left. \overbar{ \dfrac{\partial Z_2}{\partial x}} \right \vert_{1}& \left.\overbar{ \dfrac{\partial Z_2}{\partial x}}\right\vert _{2} & \cdots & \left.\overbar{ \dfrac{\partial Z_2}{\partial x}}\right\vert _{n_{\text{spot}}} & 
\left.\overbar{ \dfrac{\partial Z_2}{\partial y}}\right\vert _{1} &
\left.\overbar{ \dfrac{\partial Z_2}{\partial y}}\right\vert _{2} &
\cdots & \left.\overbar{ \dfrac{\partial Z_2}{\partial y}}\right\vert _{n_{\text{spot}}} \\
\vdots & \vdots & \ddots & \vdots & \vdots & \vdots & \ddots & \vdots \\
\left. \overbar{ \dfrac{\partial Z_J}{\partial x}} \right \vert_{1}& \left.\overbar{ \dfrac{\partial Z_J}{\partial x}}\right\vert _{2} & \cdots & \left.\overbar{ \dfrac{\partial Z_J}{\partial x}}\right\vert _{n_{\text{spot}}} & 
\left.\overbar{ \dfrac{\partial Z_J}{\partial y}}\right\vert _{1} &
\left.\overbar{ \dfrac{\partial Z_J}{\partial y}}\right\vert _{2} &
\cdots & \left.\overbar{ \dfrac{\partial Z_J}{\partial y}}\right\vert _{n_{\text{spot}}} \\
\end{bmatrix}^{T}.
\end{equation}
The system of \autoref{eq:unsolved_system} can then be written as
\begin{equation}
\mathbf{s} \approx G \cdot \mathbf{a},
\end{equation}
where $\mathbf{a}$ is the vector containing the Zernike coefficients. The least-squares estimation of $a_n^m$ becomes
\begin{equation} \label{eq:LSQ_coeff}
\mathbf{a} \approx G^{+} \cdot \mathbf{s},
\end{equation}
where $G^{+}$ the generalized inverse of the geometry matrix. This is an approximation as $G$ only contains the information of a finite number of Zernike modes, and their contribution is averaged over the lenslet array. 

\subsubsection{Method B} \label{sec:janssen_theory}
Method B relies on an analytical relation found between the local derivatives of the wavefront and Zernike polynomials. This is in contrast with method A, where there is a link between the local derivatives of the wavefront and the derivatives of the Zernike polynomials. 

The reconstruction using Method B is based on the identities:
\begin{equation}
\begin{split}
\frac{\partial Z_n^m}{\partial x} &= \frac{\partial Z_{n-2}^m}{\partial x} + n \left( Z_{n-1}^{m-1} + Z_{n-1}^{m+1} \right) \\
\frac{\partial Z_n^m}{\partial y} &= \frac{\partial Z_{n-2}^m}{\partial y} + i n \left( Z_{n-1}^{m-1} + Z_{n-1}^{m+1} \right), \\
\end{split}
\end{equation}
where these identities allow the expression of any derivative Zernike polynomial as a sum of Zernike polynomials. Here we use the convention of \Autoref{eq:complex_def} and \Autoref{eq:rmn} for the Zernike circle polynomials.

\citep{janssen2014zernike} has found that the LMS complex coefficients are given as (see appendix):
\begin{equation} \label{eq:janss_coeff}
\hat{\alpha}_n^m = C_n^m \varphi_n^m - C_{n+2}^m \varphi_{n+2}^m,
\end{equation}
where
\begin{align}
C_n^m &= \frac{1 + \delta_{n \absm}}{2n} \\ \label{eq:phinm}
\varphi_n^m &= \frac{1}{2} \left( \beta_+ \right)_{n-1}^{m+1} +  \frac{1}{2} \left( \beta_- \right)_{n-1}^{m-1},
\end{align}
and where $\delta_{nn'}$ is the Kronecker delta equal to $1$ if $n = n'$ and $0$ otherwise. Furthermore $\beta_+$ and $\beta_-$ are the Zernike coefficients of $\frac{\partial W}{\partial x} \pm i\frac{\partial W}{\partial y}$ so that
\begin{align} \label{eq:w}
 \frac{\partial W}{\partial x} \pm i\frac{\partial W}{\partial y}= \sum_{m = -\infty}^{\infty} \sum_{n \in \eta_m} \left(\beta_{\pm}\right)_n^m Z_n^m.
\end{align}

Note that the LMS coefficients $\hat\alpha_n^m$ are analytically related to four $\beta$ coefficients, namely $\left( \beta_+ \right)_{n+1}^{m+1},~\left( \beta_+ \right)_{n-1}^{m+1}, \left( \beta_- \right)_{n-1}^{m-1}$, and $\left( \beta_- \right)_{n+1}^{m-1}$. As a consequence of orthogonality of the $Z_n^m$ in the right-hand side of \Autoref{eq:w}, the LMS coefficients, once estimated, do not change anymore when the number of Zernike terms used in a finitized version of \Autoref{eq:w} is increased further.

The fact that $\hat\alpha_n^m$ is analytically related to $\beta$ coefficients is desirable, because $\beta$-coefficients can directly be estimated (in a least-squares sense) from measurable quantities. This fit is made in the same way as the least-squares method. This means that also the complex Zernike polynomials need to be averaged over the lenslets. The fit to get the $\beta$ coefficients, however, is done with a different basis than in the least-squares method to find the $a$-coefficients. The effects of this are discussed in the following section.

As a note for this method, when $n = \absm$, there will be non-existent combinations of $n$ and $m$ in \Autoref{eq:phinm}. In that case the value of $\beta$ will be set to $0$. For instance, $\hat{\alpha}_1^1$ is among others dependent on $\left(\beta_+\right)_0^2$, which goes against the constraint $|m| \le n$. To go from complex coefficient $\alpha_n^m$ to the real coefficient $a_n^m$, the relations in \Autoref{eq:alphatoa} can be used. 

In \citep{ma}, a seemingly different approach, based on vector polynomials, is used to express the wavefront Zernike coefficients in terms of wavefront slope data. However, we show in the appendix that this method is essentially equivalent to Janssen's algorithm in \citep{janssen2014zernike}.

\subsubsection{The main difference between Method A and Method B} \label{sec:cross_talk}
The main difference between Method A and Method B for finding the coefficients is in how the fitting is done. Both methods use a least-squares fit using a geometry matrix, but the matrix elements are constructed differently. In Method A, the geometry matrix elements are evaluations of the average gradient of the real-valued Zernike polynomials over certain subdisks, while for Method B it is the average of the complex-valued Zernike polynomials over the same windows. 

The gradients of the Zernike polynomials are known not to be orthogonal. This can cause problems called crosstalk when fitting the coefficients, especially when there are more aberrations present in the system than are being fit. 

If $\mathbf{a}$ is an $M$-dimensional vector containing the coefficients of the aberrations present in the system, the slopes on the Shack-Hartmann sensor can be determined as
$\mathbf{s} = G \mathbf{a}$, where $\mathbf{s}$ is an $2n_{\text{spot}}$ long vector containing the $x$- and $y$-displacement on the Shack-Hartmann sensor and $G$ an $2n_{\text{spot}} \times J$ geometry matrix defined in \Autoref{eq:geo_mat}.

The notation $Z_j=z_n^m$ is used since $j,n,m$ are related (see Eq. \Autoref{eq:j}-\Autoref{eq:m}). $Z_0$ is not included in the matrix since $Z_0=z_0^0=1$, and its partial derivatives are equal to 0. Note that the first column contains the $x$- and $y$-derivatives of the first Zernike polynomial evaluated in all $n_{\text{spot}}$ points. When a least-squares estimation of the coefficients $\hat{\mathbf{a}}$ (where the hat means to indicate that it is an estimated parameter) is made using less Zernike polynomials, up to Zernike polynomial $J<M$, crosstalk will occur. The estimator $\hat{\mathbf{a}}$ can be expressed as
\begin{equation}
\begin{align}
\hat{\mathbf{a}} &\approx G_l^{+} \mathbf{s}\\
\hat{\mathbf{a}} &\approx G_l^{+}G \mathbf{a},
\end{align}
\end{equation}
where $G_l$ is the geometry matrix containing the columns of the first $J$ Zernike polynomials. The estimator will estimate the lower-order values of the coefficients with influence of the higher-order values, because the matrix $G_l^{+} G$ will not be an identity matrix.  

When estimating the coefficient $a_n^m$, a higher-order aberration $a_{n'}^{m'}$ will influence the estimation if it is not accounted for in $G_l$ (i.e., the single index of $a_{n'}^{m'}$ $j > J$) and if
\begin{align*}
\lbrace ( n, m, n', m' ) \in \mathbb{Z}~\vert~ n \in \eta_m,&~m = m' ~\text{or}~ m = m' \pm 2, \\ & n' > n, ~n \geq m'+2,~n' \in \eta_{m'},~m' \neq 0 \rbrace,
\end{align*} 
or if 
\begin{align*}
\{ (n, m, n', m') \in \mathbb{Z} ~\vert~ n \in \eta_m,~ &m = 0 ~\text{or}~ m = 2, \\ & n' > n,~n \geq 2,~n' \in \eta_{m'},~m' = 0 \},
\end{align*}
where in both cases $\eta_m$ and $\eta_{m'}$ are the sets of allowed values for $n$ and $n'$ dependent on $m$ and $m'$ such that Equations \ref{eq:nbigger0}, \ref{eq:evennm} and \ref{eq:msmallern} are all met.

Because of this dependence of an expansion  coefficient on the maximum degree of the system of equations, it is expected that methods such as the least-squares method will incorrectly estimate the coefficients when there are higher-order aberrations present that are not accounted for in the geometry matrix $G_l$. For Method B, the geometry matrix contains the Zernike polynomials themselves, and therefore it is not expected to present any crosstalk. This is experimentally verified and shown in this paper. 
\section{Experimental setup and methods} \label{sec:setup2_method2}
\begin{figure}
\centering
\includegraphics[width = 0.9\textwidth]{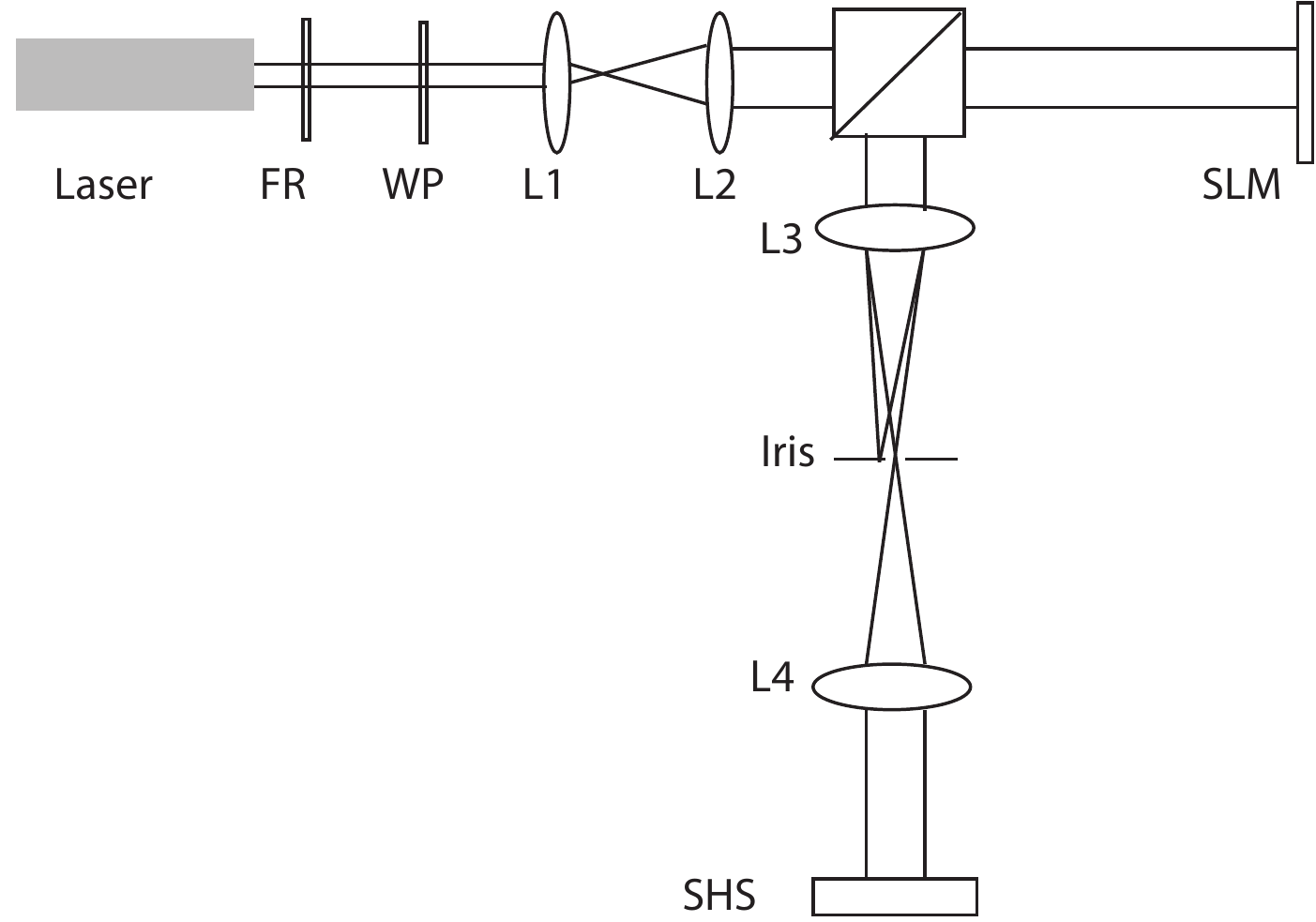}
\caption{Scheme of the experimental setup. FR is a Faraday isolator, WP is a halfwave plate, SLM is a spatial light modulator, SHS is the Shack Hartmann wavefront sensor. The zeroth order light of the SLM is blocked by the iris.\label{fig:slant_int} }
\end{figure}

In \autoref{fig:slant_int}, a schematic view of the experimental setup is shown. The phase of an uniform, collimated laser beam (HeNe laser) is modified by a spatial light modulator (Holoeye PLUTO-2-VIS-056) in such a way that known aberrations are added to the beam. In order to remove the zeroth- order reflected light from the SLM, a phase ramp is added to all SLM patterns so that the unaberrated spot is blocked by the iris after being focussed by lens L3. The lenses L1 and L2 and L3 and L4 form two 4$f$ systems.

Using the described experimental setup, all steps necessary to implement method A and Method B can be performed. These are shown in Algorithm \ref{alg:complete}. 

\begin{algorithm}
\caption{Complete measurement and comparison of Shack-Hartmann phase retrieval algorithms \label{alg:complete}}
\begin{algorithmic}[1]
\State Remove initial aberrations of the entire system 
\State Add controlled aberration $\mathbf{a}$ to the SLM 
\State Gather flat and aberrated wavefront Hartmannograms, i.e., the image of the spot pattern generated by the Shack-Hartmann sensor 
\State Find the optimal center and radius position for both methods separately such that the RMS value is minimized 
\State Calculate Zernike coefficients using optimized center and radius with a desired maximum degree
\State Calculate error between reference and estimate wavefront
\end{algorithmic}
\end{algorithm}

\subsection{Addressing a phase pattern to the SLM} \label{sec:remove_abb}
In order to be sure that only the aberration added on the SLM is measured, first all initial aberrations are eliminated. These aberrations can include alignment errors and the surface of the SLM itself, which might not be completely flat \citep{matsumoto2008generation}. Removing the initial aberrations has been done by using the Shack-Hartmann sensor and retrieving the wavefront using the least-squares method.

The way a phase pattern is addressed to the SLM is as follows. First, an aperture on the SLM is defined. In the current research this is a circular aperture, but the same method is valid if an annular aperture is used. All pixels $i$ within this aperture are used. A ``Zernike matrix'' can be set up, such that for each pixel $i$ within the aperture the value of all necessary Zernike polynomials can be computed. In matrix form this would be
\begin{equation} \label{eq:zernike_mat}
Z = \begin{bmatrix}[1.5]
\left. Z_1\right \rvert_1 & %
\left. Z_2 \right \rvert_1 & %
\cdots &%
\left. Z_J \right \rvert_1 \\
\left. Z_1\right \rvert_2 & %
\left. Z_2 \right \rvert_2 & %
\cdots &%
\left. Z_J \right \rvert_2 \\
\vdots & \vdots & \ddots & \vdots \\
\left. Z_1\right \rvert_I & %
\left. Z_2 \right \rvert_I & %
\cdots &%
\left. Z_J \right \rvert_I \\
\end{bmatrix},
\end{equation}
where $J$ is the total number of Zernike polynomials evaluated, and $I$ is the total number of pixels within the aperture. Due to the cyclic nature of the phase pattern and the limits of the SLM, the phase difference assigned to the SLM should be between 0 and $2\pi$. This can be done using the modulo (or $\bmod$) operation: $a \bmod n=a-n \left\lfloor a/n \right\rfloor$. 

If $\mathbf{p}$ is a vector containing the values of the individual pixels of the SLM, it can be constructed from the vector $\mathbf{a}$ containing the coefficients for the to-be-added aberration by
\begin{equation} 
\mathbf{p} = (Z \mathbf{a}) \bmod 2\pi.
\end{equation}
Therefore, in the case of updating the SLM pattern we use
\begin{equation}
\mathbf{p}_{new} = (\mathbf{p}_{old} - \gamma Z \mathbf{\hat{a}}) \bmod 2\pi,
\end{equation}
where $\gamma$ is the gamma function of the SLM, calibrated for the used wavelength.

After the correction is done and the wavefront from the SLM is flattened, the phase pattern of the ``flat'' phase is saved. To this phase pattern, a phase ramp is added to separate the 0$^{th}$ and 1$^{st}$ order of the SLM together with the desired aberration, in the same fashion as the correction is added. After this, it is necessary to check if the the maximum phase change over 4 pixels is not being exceeded. Such a check is necessary to avoid aliasing of the SLM-phase. In the current research, this aliasing constraint is simplified to the constraint that the difference between two neighboring pixels should not exceed $0.5\pi$. This is evaluated by letting $p_{i,j}$ be the value of the pixel located at position $i, j$ on the SLM electrode matrix. Then, first two matrices are constructed:
\begin{equation}
\begin{split}
\Delta p_x &= p_{i, j+1} - p_{i, j},\\
\Delta p_y &= p_{i + 1, j} - p_{i, j}.
\end{split}
\end{equation}
Afterward, the element-wise minimum is taken between $\Delta p$ and $2\pi - \lvert \Delta p \rvert$ for both $x$ and $y$, in order to account for the modulated phase. If any of the values of this piecewise minimum is above $0.5\pi$, it is said to break the aliasing constraint. 
\subsection{Constructing the matrix $G$}
Recalling from the theory described in the previous section, in order to recover the coefficients describing the aberrated wavefront, $\mathbf{s}$ and $G$ have to be constructed. Here, $\mathbf{s}$ relies on the two Hartmannograms, one of the flat wavefront and one of the aberrated wavefront from the SLM, and the radius $r_{\text{SH}}$ of beam hitting the Shack-Hartmann sensor. 

To construct $G$, it is necessary to average the gradients of the Zernike polynomials or their gradients. The window over which the polynomial has to be averaged can be seen as a scaled version of the lenslet, scaled so that all illuminated lenslets fit the unit disc. In order to compute the Zernike polynomials in these windows, the center position $\mathbf{c}$ and the radius $r_{\text{SH}}$ of the beam on the Shack-Hartmann sensor have to be known. The window size is estimated by the average distance between the nearest neighbor spots on the Shack-Hartmann sensor. The center position $\mathbf{c}$ follows the average position of all the midpoints of the spots on the Shack-Hartmann sensor. The radius is estimated by calculating the length between the center and the furthest spot from the center.

After this first estimation of the center position and radius is made, a better estimation can be find by optimization. For this optimization, an error has been defined that can be minimized. In this research, a root-mean-square, or RMS type error is chosen. Using the known added aberration and the measured aberration coefficients, an RMS error can be defined. To this purpose, a new Zernike matrix similar to \autoref{eq:zernike_mat} is constructed, this time with $N$ points on a grid within the unit disc. The reference phase $\mathbf{p}_{\text{ref}}$ and the recovered phase $\mathbf{p}_{\text{rec}}$ can be constructed as
\begin{equation}
\mathbf{p} = Z \mathbf{a},
\end{equation}
where for $\mathbf{p}_{\text{ref}}$, the reference vector $\mathbf{a}_{\text{ref}}$ is used, while for $\mathbf{p}_{\text{rec}}$ the estimated coefficient vector $\hat{\mathbf{a}}$ is used. Using this definition for the reference and recovered phase, the RMS error is determined as
\begin{equation}\label{eq:rms}
\varepsilon = \norm{\frac{\mathbf{p}_{\text{ref}} - \mathbf{p}_{\text{rec}}}{N}}_2,
\end{equation}
where $\norm{\mathbf{x}}_2$ is the Euclidean vector norm of vector $\mathbf{x}$. This RMS error is then minimized for center position and radius of the beam on the Hartmannogram. A limited memory bound Broyden-Fletcher-Goldfarb-Shanno (L-BFGS-B) minimization algorithm is applied to find the center position and radius\citep{fletcher}. The termination conditions for this optimization are:
\begin{align}
\frac{f^k - f^{k+1}}{\max\{|f^k|,|f^{k+1}|,1\}} &\leq 10^{-5}, \\
\max\{|\text{proj} (g_i) | ~i = 1, ..., n\} &\leq10^{-5},\\
k &\geq 10^3,
\end{align}
where $f^k$ is the value of the RMS error of the $k^{\text{th}}$ iteration of the minimization algorithm, and $\text{proj} (g_i)$ is $i^{\text{th}}$ component of the projected gradient where $n$ projections are made. If any of these statements were true, the optimization was terminated. After looking at the RMS error landscapes, it was found that not every minimum found was a global minimum. If this was the case, the global minimum coordinates were estimated using the RMS landscape graphs, and a brute force optimization was run around those coordinates. This brute force optimization calculates the RMS error value in a grid of points. From the coordinates with the lowest RMS error, a new downhill simplex minimization algorithm is started. This way the global minimum was attempted to be found, and the optimal center and radius positions were determined.

In the current research, the necessary aberration coefficients are obtained using a set of Zernike polynomials with a maximum degree of eight. From \citet{soloviev2006estimation} it is concluded that a good rule of thumb for the amount of Zernike polynomials that can be fit given $k$ spots on the Shack-Hartmann sensor is $\sfrac{k}{3}$. In the current research, there were about 144 spots on the Shack-Hartmann sensor, and there are 45 polynomials in the Zernike expansion with a maximum degree of eight. 

For method A, the amount of polynomials fit is equal to the amount of retrieved coefficients, while for Method B, it is not so, since for the latter if a fit is made with up to maximum degree of eight, the first seven degrees can be retrieved. 

It should be noted that this optimization can take long due to the fact that the geometry matrix needs to be calculated in every iteration, as the values in the matrix depend on the center position and radius. 

These optimized parameters can then be used to find the RMS value fitting any degree of Zernike polynomials, and can also be used to determine the error landscape by calculating the RMS values when the center and radius differ slightly from the optimal value.  

\section{Results} \label{sec:results}
The two methods have been implemented for three different test cases: a) single Zernike aberrations, b) a combination of three aberrations, assumed not to show crosstalk, and c) four cases of aberrations where crosstalk is present. The aberrations are listed in \autoref{tab:spec_coeff}.

\begin{table}[hbt]
\centering
\caption{Coefficients used in specific Zernike experiments, rounded off to 3 significant numbers \label{tab:spec_coeff}}
\begin{tabular}{l | r r r r r r r r r}
Code & \multicolumn{9}{l}{Coefficients}\\
\hline
\multirow{2}{*}{\texttt{5\_1}}& $a_{5}^{1}$  &  &  &  &  &  &  &  & \\ 
& 0.750  &  &  &  &  &  &  &  & \\ 
\hline\multirow{2}{*}{\texttt{5\_5}}& $a_{5}^{5}$  &  &  &  &  &  &  &  & \\ 
& 1.250  &  &  &  &  &  &  &  & \\ 
\hline\multirow{2}{*}{\texttt{6\_2}}& $a_{6}^{2}$  &  &  &  &  &  &  &  & \\ 
& 0.500  &  &  &  &  &  &  &  & \\ 
\hline\multirow{2}{*}{\texttt{6\_4}}& $a_{6}^{4}$  &  &  &  &  &  &  &  & \\ 
& 1.000  &  &  &  &  &  &  &  & \\ 
\hline\multirow{2}{*}{\texttt{6\_6}}& $a_{6}^{6}$  &  &  &  &  &  &  &  & \\ 
& 2.000  &  &  &  &  &  &  &  & \\ 
\hline\multirow{2}{*}{\texttt{3\_zerns\_1}}& $a_{2}^{0}$ & $a_{4}^{4}$ & $a_{6}^{-2}$  &  &  &  &  &  & \\ 
& 2.000 & 1.500 & 0.500  &  &  &  &  &  & \\ 
\hline\multirow{2}{*}{\texttt{3\_zerns\_2}}& $a_{2}^{0}$ & $a_{3}^{-1}$ & $a_{5}^{5}$  &  &  &  &  &  & \\ 
& 2.000 & 3.000 & 1.500  &  &  &  &  &  & \\ 
\hline\multirow{2}{*}{\texttt{3\_zerns\_3}}& $a_{2}^{-2}$ & $a_{4}^{0}$ & $a_{6}^{-4}$  &  &  &  &  &  & \\ 
& 2.000 & 1.500 & 0.750  &  &  &  &  &  & \\ 
\hline\multirow{2}{*}{\texttt{sub\_zerns\_1}}& $a_{2}^{0}$ & $a_{4}^{0}$ & $a_{6}^{0}$  &  &  &  &  &  & \\ 
& 4.000 & 1.500 & 0.750  &  &  &  &  &  & \\ 
\hline\multirow{2}{*}{\texttt{sub\_zerns\_2}}& $a_{1}^{1}$ & $a_{3}^{1}$ & $a_{5}^{1}$  &  &  &  &  &  & \\ 
& 4.000 & 1.500 & 0.750  &  &  &  &  &  & \\ 
\hline\multirow{2}{*}{\texttt{sub\_zerns\_3}}& $a_{3}^{3}$ & $a_{5}^{3}$  &  &  &  &  &  &  & \\ 
& 2.500 & 0.750  &  &  &  &  &  &  & \\ 
\hline\multirow{2}{*}{\texttt{sub\_zerns\_4}}& $a_{2}^{2}$ & $a_{4}^{2}$ & $a_{6}^{2}$  &  &  &  &  &  & \\ 
& 2.500 & 1.000 & 0.500  &  &  &  &  &  & \\ 
\hline
\end{tabular}
\end{table}

\FloatBarrier

\begin{figure}
\centering
\includegraphics[width = \textwidth]{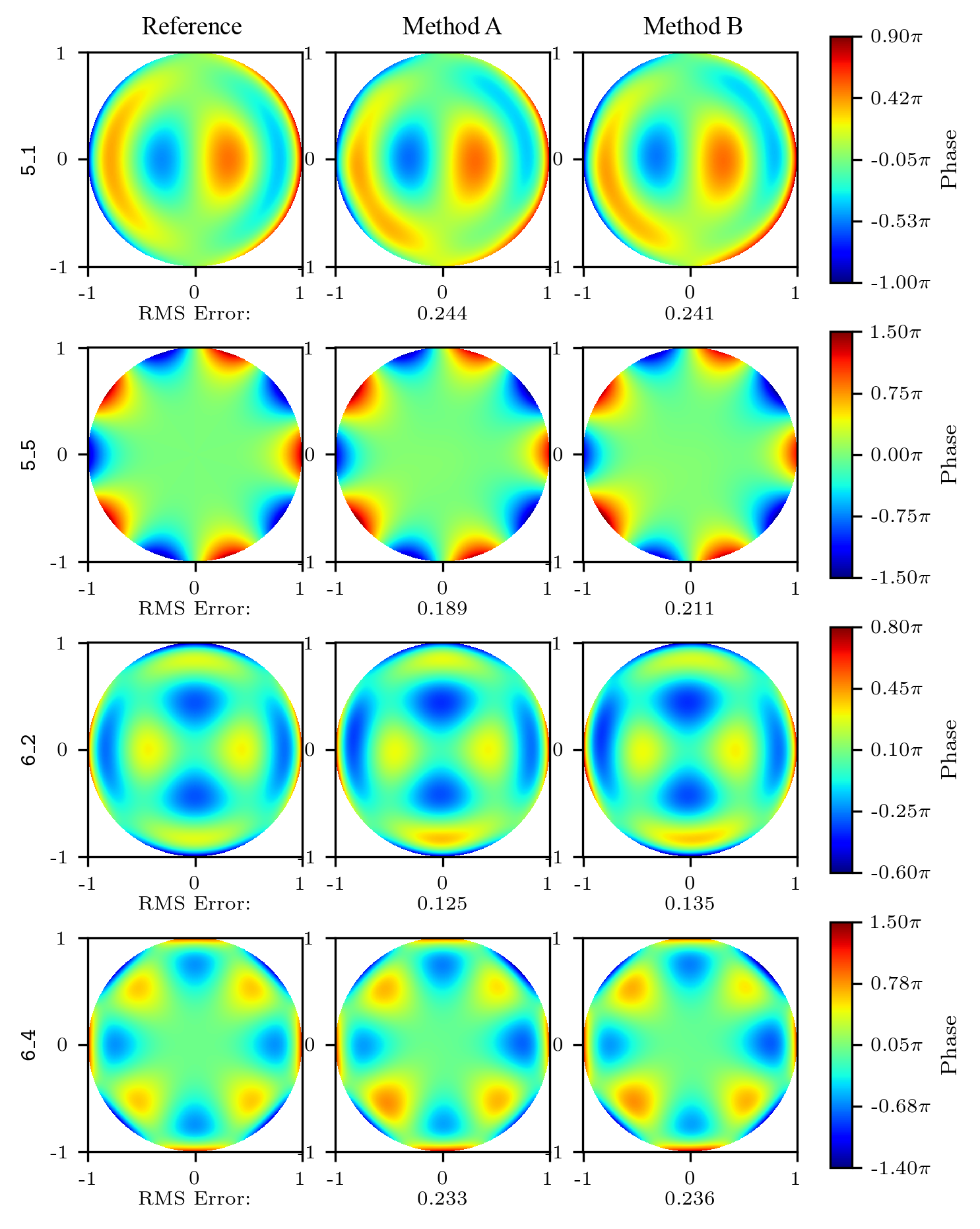}
\caption{The first 4 added aberrations (left column) and their reconstructions (middle and right columns). RMS errors have been indicated below the reconstructions. \label{fig:recon_504_0}}
\end{figure}

\begin{figure}
\centering
\includegraphics[width = \textwidth]{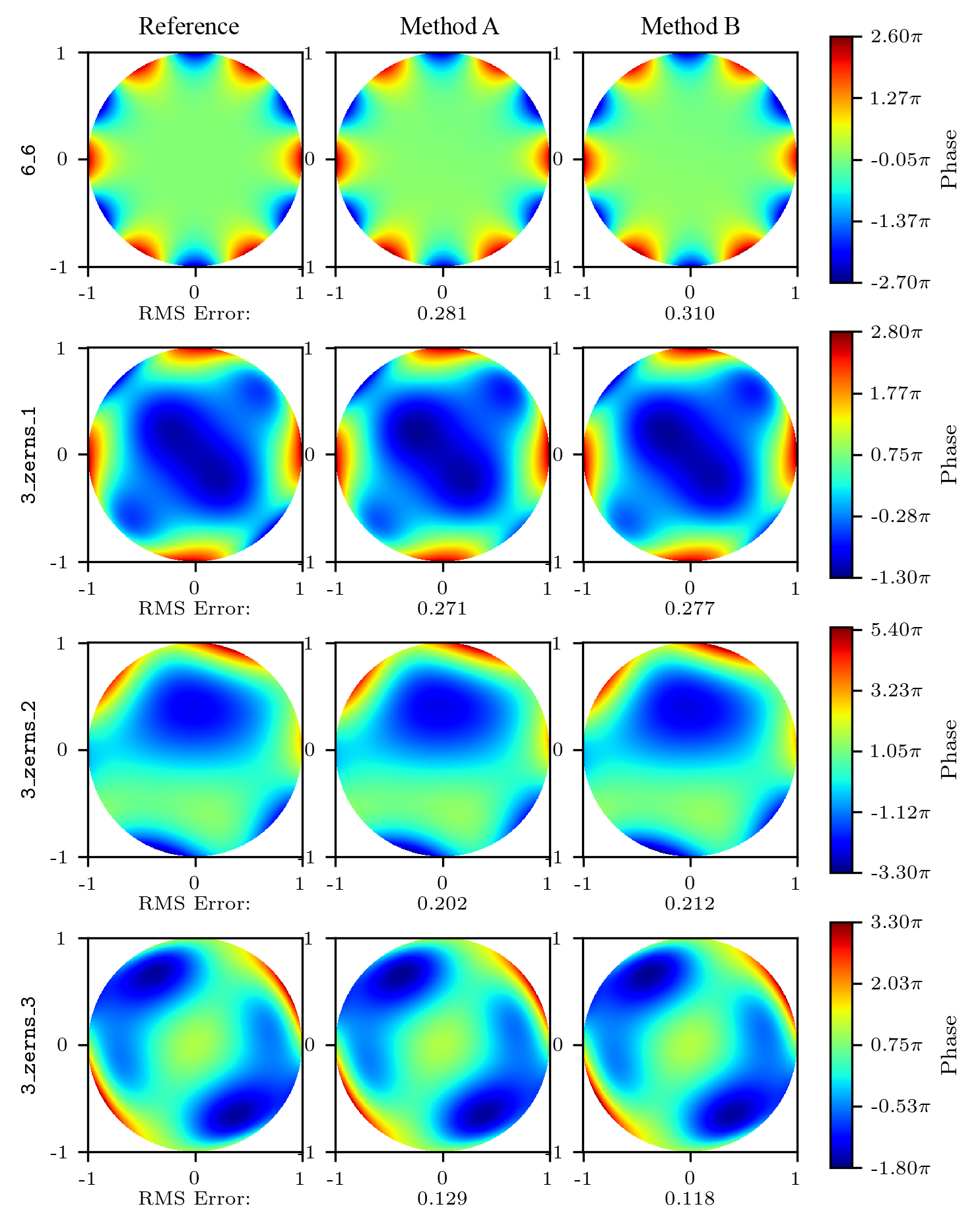}
\caption{The second 4 added aberrations (left column) and their reconstructions (middle and right columns). RMS errors have been indicated below the reconstructions. \label{fig:recon_504_1}}
\end{figure}

\begin{figure}
\centering
\includegraphics[width = \textwidth]{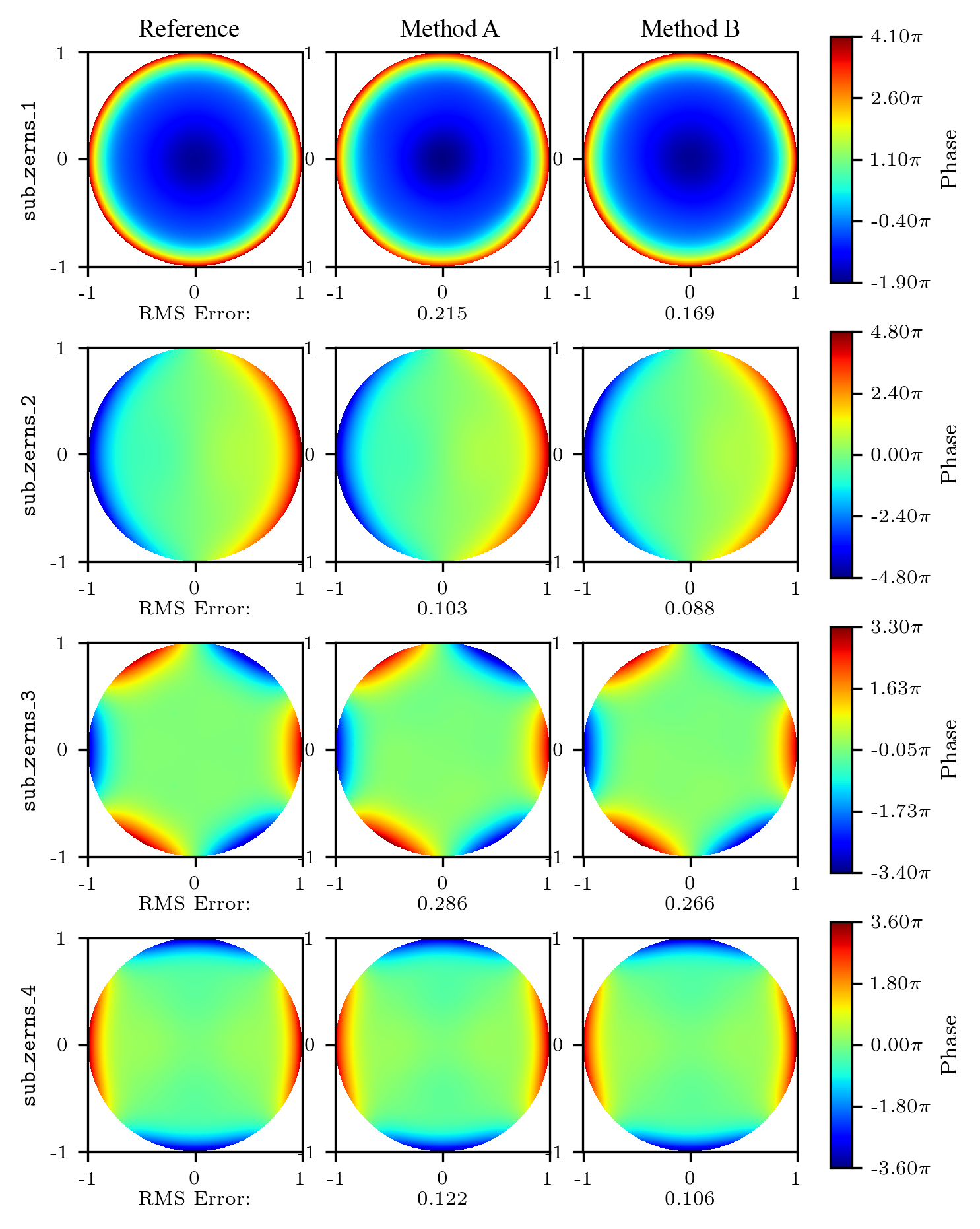}
\caption{The last 4 added aberrations (left column) and their reconstructions (middle and right columns). RMS errors have been indicated the reconstructions. \label{fig:recon_504_2}}
\end{figure}

The reconstructions using the recovered coefficients are shown in \autoref{fig:recon_504_0} - \autoref{fig:recon_504_2}. The left column shows the added aberration on the SLM, while the middle and right column show the wavefront reconstruction using our own implementation of the least-squares method (Method A) and Method B, respectively. The RMS error is indicated in the figures. From these results, it can be seen that the Method B compares very well with Method A and the input wavefront.

\FloatBarrier

Based on the errors in the reconstruction found in  \autoref{fig:recon_504_0}, \autoref{fig:recon_504_1} and \autoref{fig:recon_504_2}, one can see that both methods provide comparable accuracy in reconstructing the wavefront. However, it is known that Method A presents crosstalk of coefficients when less Zernike powers are fit than there are aberrations present in the system. In the following, we show experimentally that this is not the case for Method B.

Using the same center position and radius, different amount of Zernike degrees can be fit in order to see the convergence behavior of gathered coefficients. Due to the fact that there are only 1, 2, or 3 Zernike modes present in the specific Zernike experiment, the convergence behavior of these experiments can be visualized.

\begin{figure}
\centering
\includegraphics[width = \textwidth]{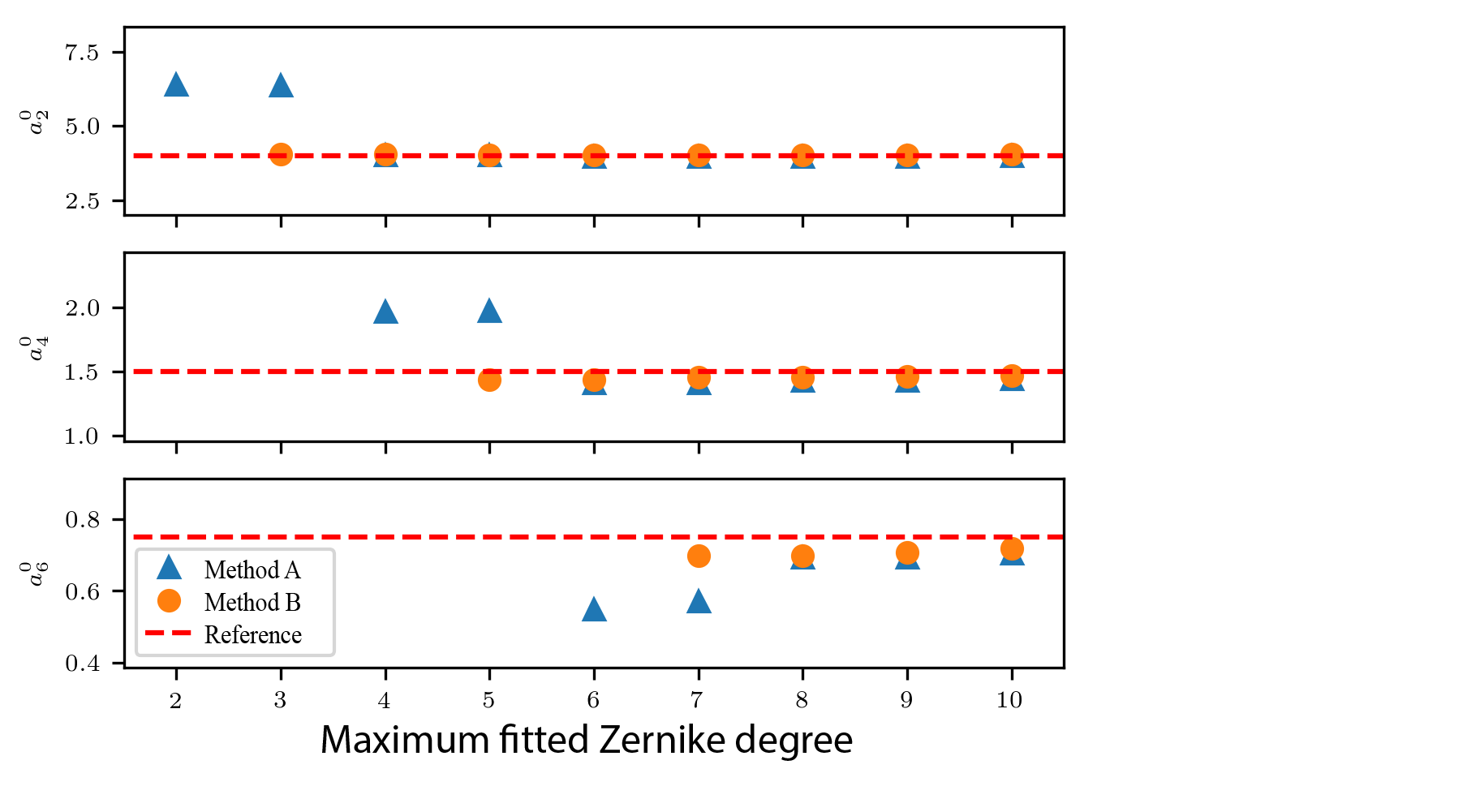}
\caption{Convergence of coefficients for the \texttt{sub\_zerns\_1} experiment. \label{fig:coef_conv_sz1}}
\end{figure}

\begin{figure}
\centering
\includegraphics[width = \textwidth]{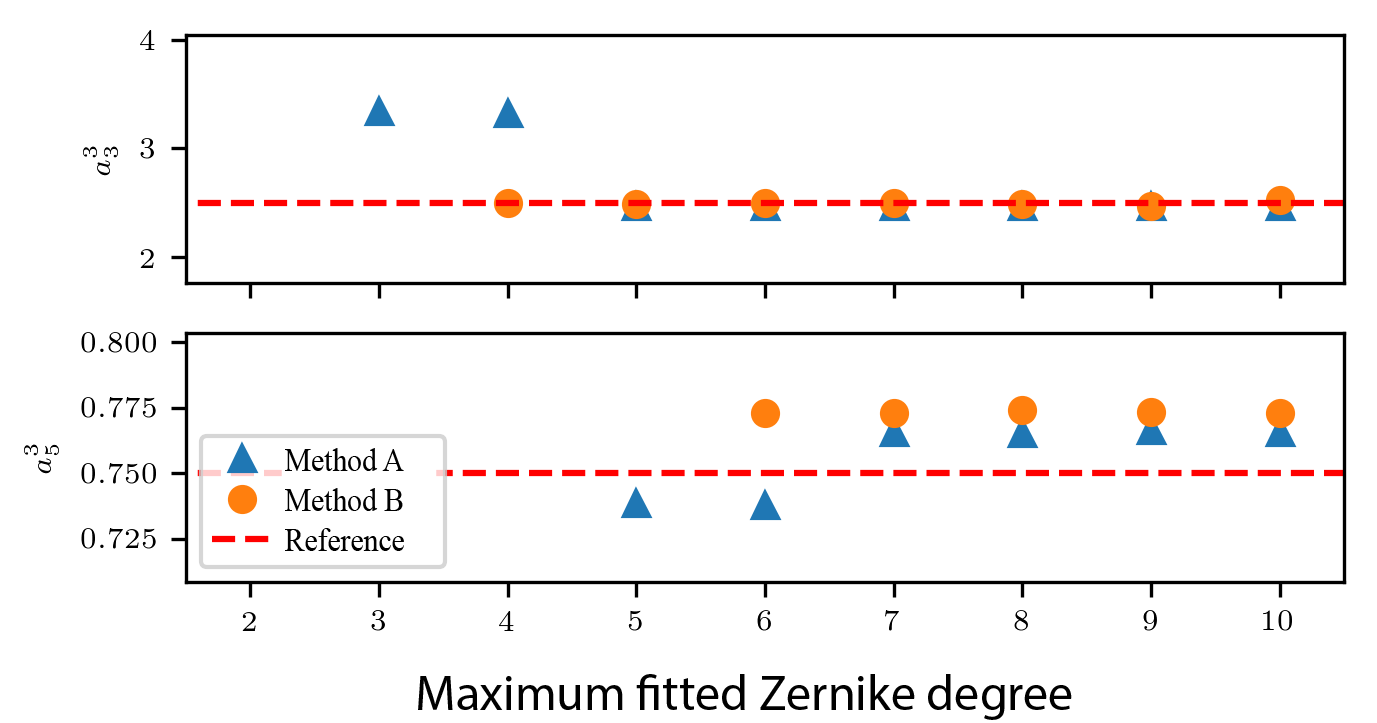}
\caption{Convergence of coefficients for the \texttt{sub\_zerns\_3} experiment. \label{fig:coef_conv_sz3}}
\end{figure}

Based on analysis of the aberrations listed in \autoref{tab:spec_coeff}, the single Zernike experiments \verb|5_1| and \verb|6_4| and 3 random Zernike experiments \verb|3_zerns_1| and \verb|3_zerns_3|, one can see in \autoref{fig:recon_504_0} and \autoref{fig:recon_504_1} that the coefficients seem to be measured independently from each other, and there is no difference between the two methods. However, looking at the subsequent Zernike experiments \verb|sub_zerns_1| and \verb|sub_zerns_3| a difference can be seen between the two reconstruction methods. The initial guesses of  Method A over- or underestimate the presence of the aberration when the fit is performed with a maximum degree of eight. From \autoref{fig:coef_conv_sz1} it can be seen that defocus is overestimated by more than 50\% until 4 degrees are fit. At maximum degree of four, the spherical aberration coefficient $a_4^0$ is overestimated. All values seem to be within normal range when fitting a maximum degree of eight. The same can be seen in \autoref{fig:coef_conv_sz3}, where the coefficient $a_3^3$ is overestimated at maximum degrees of three and four. Method B does not present this over-estimation.

\section{Conclusion and outlook} \label{sec:conclusion}
In conclusion, we have implemented a new wavefront reconstruction method (Method B) for Shack-Hartmann sensors based on Zernike expansion of derivatives of the Zernike circle polynomials that was introduced in \citep{janssen2014zernike}. We have shown with experiments that Method B is advantageous as compared to other known methods such as a least-squares method (Method A) due to the lack of cross-talk between the coefficients. We have also compared Method B to Method A for reconstructing the wavefront using a Shack-Hartmann sensor for some sets of aberrations. 

Based on the error values of the reconstruction with optimal amount of Zernike coefficients, the quality of the fit is in general similar for Method A and Method B. When less than the optimal Zernike powers were fit, it was seen that Method B estimates the coefficients more accurately. For the single Zernike aberration case, it was shown that only Method A shows cross-coupling of higher order aberrations, while Method B does not. 

As to computational load, we may point out that these are of comparable order, with Method B somewhat more demanding due to the complex arithmetics and the alignment needed for compatibility with the ANSI-format of the Shack-Hartmann sensor.

Finally, we would like to point out that there are also alternative expansions to treat this problem, namely the use of eigenfunctions of the Laplacian with Neumann boundary conditions\citep{huang1},\citep{huang2} that do have orthogonal gradients. It would be interesting as future work to compare this method with Method A and Method B considered in the present paper.

\bibliography{bib3} \clearpage
\appendix
\section{Relation with orthogonal vector polynomials}
\label{app:proof}
In the course of our investigations, we became aware of a seemingly different approach to obtain the wavefront aberration coefficients from wavefront derivative data. We shall describe and relate this approach to our method in the framework and notations of \citep{janssen2014zernike}, so that the Zernike circle polynomials are unnormalized and have exponential azimuthal dependence facilitating mathematical developments. The reader will have no particular problems in reformulating the main results of this appendix in terms of the ANSI-style circle polynomials using (Eqs. 2.14$-$16). Thus, we have for integer $n$ and $m$ such that $n-|m|$ is even and non-negative:
\begin{equation}
\label{eq:x1}
Z_{n}^{m}(\nu,\mu)\equiv Z_{n}^{m}(\rho,\theta)=R_{n}^{|m|}(\rho)e^{im\theta},
\end{equation}
with real $\nu,\mu$ such that $\nu^2+\mu^2 \le 1$ and
\begin{equation}
\label{eq:x2}
\nu+i\mu=\rho e^{i\theta}; \ \nu=\rho\cos\theta; \ \mu=\rho \sin\theta,
\end{equation}
and the radial polynomials $R_{n}^{|m|}(\rho)$ given by Eq. 2.11. 

We now sketch the approach in \citep{ma}, where the notations and conventions differ from our present analysis. The approach \citep{ma} uses the notion of vector polynomials
\begin{equation}
\label{eq:x3}
\underline{G}_n^m=\underline{G}_n^m(\rho, \theta)=(G_{n,1}^{m}(\rho,\theta),G_{n,2}^{m}(\rho,\theta)) \in \mathbb{C}^2
\end{equation}
for integer $n,m$ such that $n-|m|$ is even and non-negative with $n\ne0$, that satisfy
\begin{equation}
\label{eq:x4}
\frac{1}{\pi}\int_0^{1}\int_0^{2\pi} \nabla Z_n^m(\rho,\theta).\underline{G}_{n'}^{m'*}(\rho,\theta)\rho d\rho d\theta=\delta_{m,m'}\delta_{n,n'}
\end{equation}
for integer $n,m,n',m'$ such that $n-|m|$ and $n'-|m'|$ are even and non-negative while $n\ne 0 \ne n'$. In \Autoref{eq:x4} we denote for $(z,w)\in \mathbb{C}^2$ and $(g,h)\in \mathbb{C}^2$
\begin{equation}
\label{eq:x5}
(z,w) \cdot (g,h)=zg+wh,	
\end{equation}
the $*$ denotes complex conjugation, and
\begin{equation}
\label{eq:x6}
(\nabla Z_n^m)(\nu,\mu)=\left(\frac{\partial Z_n^m}{\partial \nu}(\nu,\mu),	\frac{\partial Z_n^m}{\partial \mu}(\nu,\mu)\right) \in \mathbb{C}^2.
\end{equation}
With $\alpha_n^m$ the Zernike coefficients of $W$ (to be found) according to Eq. 2.13,
where, due to orthogonality (see Eq. 2.12),
\begin{equation}
\label{eq:x8}
\alpha_n^m=\frac{n+1}{\pi}\int_0^1\int_0^{2\pi} W(\rho,\theta) Z_n^{m*}(\rho,\theta)\rho d\rho d\theta,
\end{equation}
we have
\begin{equation}
\label{eq:x9}
\nabla W=\sum_{n,m}\alpha_n^m\nabla Z_n^m,
\end{equation}
and
\begin{equation}
\label{eq:x10}
\alpha_n^m=\frac{1}{\pi}\int_0^1\int_0^{2\pi} \nabla W(\rho,\theta) \underline{G}_n^{m*}(\rho,\theta)\rho d\rho d\theta.
\end{equation}
In \citep{ma}, sect.2A, the condition \Autoref{eq:x4} is elaborated using Green's theorem, where it should be noted that in \citep{ma} the attention is restricted to the cases $n=n',m=m'$ in \Autoref{eq:x4}. It is shown that the problem of satisfying \Autoref{eq:x4}, with $n=n',$ $m=m'$ is solved when we can find $\underline{G}_n^m$ such that on the pupil
 \begin{equation}
\label{eq:x11}
\nabla. \underline{G}_n^m=\frac{\partial G_{n,1}^m}{\partial \nu}+	\frac{\partial G_{n,2}^m}{\partial \mu} =-(n+1)Z_n^m,
\end{equation}
while the boundary condition
 \begin{equation}
\label{eq:x12}
G_{n,1}^m (\rho=1,\theta)\cos\theta+ G_{n,2}^m(\rho=1,\theta)\sin\theta=0
\end{equation}
should be satisfied at the rim $\rho^2=\nu^2+\mu^2=1$ of the pupil. 
 Next, in \citep{ma}, subsect. 2.B.1, the $\underline{G}_n^m$ are required to be irrotational, meaning that the $\nabla \times \underline{G}_n^m=0$, because of considerations of minimal noise propagation. This condition of irrotationality is satisfied when there is a scalar function $U_n^m$ on the pupil such that
\begin{equation}
\label{eq:x13}
\underline{G}_n^m=\nabla U_n^m.
\end{equation}
Combining \Autoref{eq:x13} and \Autoref{eq:x11}, we see that we want $U_n^m$ to satisfy
\begin{equation}
\label{eq:x14}
\nabla^2 U_n^m=\Delta U_n^m=-(n+1) Z_n^m
\end{equation}
while the boundary condition \Autoref{eq:x12}
\begin{equation}
\label{eq:x15}
\frac{\partial U_n^m}{\partial \nu}(\rho=1,\theta)\cos\theta+ \frac{\partial U_n^m}{\partial \mu}(\rho=1,\theta)\sin\theta=0
\end{equation}
should hold on the rim $\rho=1$ of the pupil. In \citep{ma} subsect. 2.B.2-3, the $U_n^m$ are found by writing the condition \Autoref{eq:x14} in polar coordinates, with separate consideration of the cases $m=0$ and $m\ne0$, and explicitly using the series representation of radial polynomials. This yields the two components of $G_n^m$ in trigonometric polynomial form, for which it can be shown that the boundary condition 
\Autoref{eq:x12} is satisfied as well. In \citep{janssen2014zernike}, sect. 4, it is shown that \Autoref{eq:x14} has a solution
\begin{equation}
\label{eq:x16}
U_n^m=-\left[\frac{Z_{n+2}^m}{4(n+2)}-\frac{(n+1)Z_{n}^m}{2n(n+2)}+\frac{Z_{n-2}^m}{4n}\right].
\end{equation}
We shall verify below that, when $n=|m|+2,|m|+4,...$, this $U_n^m$ also satisfies the boundary condition \Autoref{eq:x15}, and that a concise formula for $\underline{G}_n^m=\nabla U_n^m$ in terms of Zernike circle polynomials results. For the case that $n=|m|$, we shall show that
\begin{equation}
\label{eq:x17}
U_n^m=-\left[\frac{Z_{n+2}^m}{4(n+2)}-\frac{(3n+4)Z_{n}^m}{4n(n+2)}\right], n=|m|,
\end{equation}
satisfies \Autoref{eq:x14} and \Autoref{eq:x15}, and we find a concise formula in terms of the Zernike circle polynomials for $\underline{G}_{|m|}^m$ as well. This then can be used to show that $\alpha_n^m$ of \Autoref{eq:x10} actually coincide with the LMS estimator found in \citep{janssen2014zernike}, sect. 3. The trigonometric$/$polynomial solution form of $\underline{G}_n^m$ found in \citep{ma} involves the coefficients in the series representation of $R_n^{|m|}(\rho)$ in \Autoref{eq:x17}, and these become awkward to use when the degree becomes large ($n$ should be limited to $\le 44$ when using double precision). This problem is virtually absent when the representation of the $\underline{G}_n^m$ in terms of the Zernike circle polynomials is used since there are nowadays several methods for reliably computing (the radial parts of the) Zernike circle polynomials of arbitrary large degree $n$ and azimuthal order $m$, \citep{JD},\citep{SP}.

We shall now show for $m=1,2,...$ and $n=m+2, m+4, ....$ that $U_n^m$ of \Autoref{eq:x16} satisfies \Autoref{eq:x15}. Recalling the convention in \citep{janssen2014zernike} that any $Z_{n'}^{m'}$ with $|m'|>n'$ is set to 0, we have 
\begin{equation}
\label{eq:x18}
\left( \frac{\partial}{\partial \nu} \pm i\frac{\partial}{\partial \mu} \right) Z_{n'}^{m'}(\nu,\mu)=2\sum_{l=0}^{\frac{n'-|m'|}{2}} (n'-2l) Z_{n'-1-2l}^{m'\pm 1},
\end{equation}
see \citep{janssen2014zernike}, Eq. 13. Using \Autoref{eq:x18} with $\nu+i\mu=e^{i \theta}$, noting that
\begin{equation}
\label{eq:x19}
Z_{n-1-2l}^{m \pm 1}(1,\theta)=e^{i(m \pm 1)\theta}, \,l=0, 1, ... \frac{1}{2}(n-|m|-1),
\end{equation}
while (as $m>0$)
\begin{equation}
\label{eq:x20}
Z_{m-1}^{m- 1}(1,\theta)=e^{i(m-1)\theta}, \, Z_{m-1}^{m+1}(1,\theta)=0,
\end{equation}
we have
\begin{equation}
\label{eq:x21}
\left(\frac{\partial}{\partial \nu}+i \frac{\partial}{\partial \mu}\right)Z_{n}^{m}=2\sum_{l=0}^{\frac{n-m}{2}} (n-2l)e^{i(m+1)\theta}-2me^{i(m+1)\theta},
\end{equation}
\begin{equation}
\label{eq:x22}
\left(\frac{\partial}{\partial \nu}- i\frac{\partial}{\partial \mu}\right)Z_{n}^{m}=2\sum_{l=0}^{\frac{n-m}{2}} (n-2l)e^{i(m-1)\theta}.
\end{equation}
Adding and subtracting \Autoref{eq:x21} and \Autoref{eq:x22} from one another then gives
\begin{equation}
\label{eq:x23}
\frac{\partial Z_n^m}{\partial \nu}=2e^{i m\theta}\cos\theta\sum_{l=0}^{\frac{n-m}{2}} (n-2l)-m e^{i(m+1)\theta},
\end{equation}
\begin{equation}
\label{eq:x24}
\frac{\partial Z_n^m}{\partial \mu}=2e^{i m\theta}\sin \theta \sum_{l=0}^{\frac{n-m}{2}} (n-2l)-\frac{1}{i}m e^{i(m+1)\theta}.
\end{equation}
We also observe that
\begin{equation}
\label{eq:x25}
\sum_{l=0}^{\frac{n-m}{2}}(n-2l)=\frac{n+m}{2} \left( \frac{n+m}{2}+1 \right).
\end{equation}
Then, from \Autoref{eq:x16}, \Autoref{eq:x23} and \Autoref{eq:x25}, we get
\begin{align}
\label{eq:x26}
&\frac{\partial U_n^m}{\partial \nu}(\rho=1,\theta)= \nonumber\\
&=-\left[\frac{1}{4(n+2)}\frac{\partial Z_{n}^m}{\partial \nu}-\frac{n+1}{2n(n+2)}\frac{\partial Z_{n}^m}{\partial \nu}+ \frac{1}{4n}\frac{\partial Z_{n-2}^m}{\partial \nu}\right] \nonumber\\ 
&=-2D_n^m e^{i m\theta}\cos\theta+m E_n e^{i(m+1)\theta}, 
\end{align}
where
\begin{equation}
\label{eq:x27}
D_n^m=\frac{\frac{n+2+m}{2}(\frac{n+2-m}{2}+1)}{4(n+2)}-\frac{(n+1)\frac{n+m}{2}(\frac{n-m}{2}+1)}{2n(n+2)}+\frac{\frac{n-2+m}{2}(\frac{n-2-m}{2}+1)}{4n}=0, 
\end{equation}
and
\begin{equation}
\label{eq:x28}
E_n=\frac{1}{4(n+2)}-\frac{n+1}{2n(n+2)}+\frac{1}{4n}=0.
\end{equation}
Hence, $(\partial U_n^m/\partial \nu) (\rho,\theta)=0$ for $\rho=1$ and, similarly  $(\partial U_n^m/\partial \mu) (\rho,\theta)=0$ for $\rho=1$, and this implies that \Autoref{eq:x15} holds. This handles the case that $m = 1,2,...$ and $n = m+2,m+4,...$. The case the that $m=-1, -2,...$ and $n=|m|+2, |m|+4,...$ follows from the case already handled by using that
\begin{equation}
\label{eq:x29}
Z_{n'}^{m'}(\nu,\mu)= (Z_{n'}^{-m'}(\nu,\mu))^*.
\end{equation}
The case that $m=0$ and $n=2,4,...$ can be proved in the same way as the case $m=1,2,...$ and $n=m+2, m+4,...$, where now in \Autoref{eq:x23} and \Autoref{eq:x24} the terms involving $e^{i(m+1)\theta}$ disappear.

We shall next show that for $n=|m|+2,|m|+4,...$ 
\begin{equation}
\label{eq:x30}
\frac{\partial U_n^m}{\partial \nu}=-\frac{1}{4}\left[Z_{n+1}^{m+1}+Z_{n+1}^{m-1}-Z_{n-1}^{m+1}-Z_{n-1}^{m-1}\right],
\end{equation}
\begin{equation}
\label{eq:x31}
\frac{\partial U_n^m}{\partial \mu}=-\frac{1}{4i}\left[Z_{n+1}^{m+1}-Z_{n+1}^{m-1}-Z_{n-1}^{m+1}+Z_{n-1}^{m-1}\right].
\end{equation} 
Then \Autoref{eq:x30} and \Autoref{eq:x31} yield the announced concise expression in terms of the Zernike circle polynomials for
\begin{equation}
\label{eq:x32}
\underline{G}_n^m=\left(\frac{\partial U_n^m}{\partial \nu},\frac{\partial U_n^m}{\partial \mu}\right).
\end{equation}
Using \Autoref{eq:x18} and \Autoref{eq:x16}, we have that
\begin{align}
\label{eq:x33}
\frac{\partial U_n^m}{\partial \nu} & =-[ \frac{1}{4(n+2)}\sum_{l=0}^{\frac{n+2-|m|}{2}}
(n+2-2l)(Z_{n+1-2l}^{m+1}+Z_{n+1-2l}^{m-1}) \nonumber \\
&-\frac{n+1}{2n(n+2)}\sum_{l=0}^{\frac{n-|m|}{2}}(n-2l)(Z_{n-1-2l}^{m+1}+Z_{n-1-2l}^{m-1}) \nonumber \\
&+\frac{1}{4n}\sum_{l=0}^{\frac{n-2-|m|}{2}}(n-2-2l)(Z_{n-3-2l}^{m+1}+Z_{n-3-2l}^{m-1}) ]
\end{align}
We now observe that the three series in \Autoref{eq:x33} have the same terms, except that the second series omits the term $l=0, 1$ from the first series and the third series omits the terms with $l=0$ from the first series. Using Eq. (A.28), we see that the terms in the first series with $l=2,3,...(n+2-|m|)/2$ are canceled, and we get
\begin{align}
\label{eq:x34}
\frac{\partial U_n^m}{\partial \nu} & =-[ \frac{1}{4(n+2)}((n+2)(Z_{n+1}^{m+1}+Z_{n+1}^{m-1})+n(Z_{n-1}^{m+1}+Z_{n-1}^{m-1})) \nonumber \\
&-\frac{n+1}{2n(n+2)} n (Z_{n-1}^{m+1}+Z_{n-1}^{m-1}) ] \nonumber \\
&=-\left[ \frac{1}{4}(Z_{n+1}^{m+1}+Z_{n-1}^{m+1})-\frac{1}{4}(Z_{n-1}^{m+1}+Z_{n-1}^{m-1}) \right],
\end{align} 
and this is \Autoref{eq:x30}. In a similar fashion, we get \Autoref{eq:x31}.

We now consider the case that $n=m= 1, 2,...$. Then $Z_{n-2}^m= 0$ and $\Delta Z_n^m=0$, and so it follows from \Autoref{eq:x23} that $\Delta U=-(m+1)Z_m^m$ holds for any $U$ of the form
\begin{equation}
\label{eq:x35}
U=-\left[ \frac{Z_{m+2}^m}{4(m+2)}-C Z_m^m \right].
\end{equation}
We shall determine $C$ such that \Autoref{eq:x15} holds for $n=m$. Now
\begin{equation}
\label{eq:x36}
Z_{m}^{m}(\rho,\theta)=\rho^{m} e^{i m \theta}, \,\,\, Z_{m+2}^{m}(\rho,\theta)=((m+2)\rho^{m+2}-(m+1)\rho^m)e^{i m\theta},
\end{equation}
and so the condition \Autoref{eq:x15} for $n=m$ and $\theta=0$ yields
\begin{equation}
\label{eq:x37}
C=\frac{3m+4}{4m(m+2)}.
\end{equation}
It can be shown that with this value of $C$ the $U$ of \Autoref{eq:x35} satisfies \Autoref{eq:x15} also for $\theta \neq 0$. This handles the case that $n=m= 1, 2,....$. The case $n=m=0$ is non-existent, and the case with $n=-m= 1,2,...$ follows from the case already handled by complex conjugation. We now also compute for this  $U$, using \Autoref{eq:x18}, and $m>0$
\begin{align}
\label{eq:x38}
\frac{\partial U}{\partial \nu}&= -\left[\frac{1}{4(m+2)}\frac{\partial Z_{m+2}^{m}}{\partial \nu}-\frac{3m+4}{4m(m+2)}\frac{\partial Z_{m}^{m}}{\partial \nu}\right] \nonumber \\
&=-\left[\frac{1}{4}Z_{m+1}^{m+1}+\frac{1}{4}Z_{m+1}^{m-1}+\frac{m}{4(m+2)}Z_{m-1}^{m-1}-\frac{3m+4}{4m(m+2)}mZ_{m-1}^{m-1}\right] \nonumber \\
& =-\left[\frac{1}{4}Z_{m+1}^{m+1}+\frac{1}{4}Z_{m+1}^{m-1}-\frac{1}{2}Z_{m-1}^{m-1}\right],
\end{align}
and similarly
\begin{equation}
\label{eq:x39}
\frac{\partial U}{\partial \mu}= -\frac{1}{i} \left[ \frac{1}{4}Z_{m+1}^{m+1}-\frac{1}{4}Z_{m+1}^{m-1}+\frac{1}{2}Z_{m-1}^{m-1} \right].
\end{equation}
The results \Autoref{eq:x38} and \Autoref{eq:x39} continue to hold when $m<0$ and $n=|m|$. Hence, also in this case we get a concise result for $\underline{G}_{|m|}^m$ as in \Autoref{eq:x30} $-$ \Autoref{eq:x32}.

We finally show that the $\alpha_n^m$ obtained in \citep{janssen2014zernike}, sect. 3 coincide with the $\alpha_n^m$ of \Autoref{eq:x10}. Observe that the vector polynomials $\underline{G}_m^m$ in \Autoref{eq:x10} were derived in \citep{ma} from the conditions \Autoref{eq:x4} and \Autoref{eq:x12} under the assumption that they have vanishing curls, whereas the LMS estimator found in \citep{janssen2014zernike}, sect. 3 has been derived under the condition that a natural mean-square error functional involving expansion coefficients is minimized. This LMS estimate uses the expansion coefficients $(\beta_{\pm})_n^m$ in
\begin{equation}
\label{eq:x40}
\frac{\partial W}{\partial \nu}\pm i \frac{\partial W}{\partial \mu}=\sum_{n,m}{\beta_{\pm}}_n^m Z_n^m.
\end{equation}
That is, we have for $m \ne 0$
\begin{equation}
\label{eq:x41}
\alpha_n^m = C_n^m \varphi_n^m-C_{n+2}^m \varphi_{n+2}^m, n=|m|,|m|+2, ...,
\end{equation}
with
\begin{align}
\label{eq:x42}
\varphi_n^m=&\frac{1}{2}(\beta_+)_{n+1}^{m+1}+(\beta_-)_{n-1}^{m-1}, \\
C_{|m|}^m=& \frac{1}{|m|}, \, C_n^m=\frac{1}{2n}, n=|m|+2,|m|+4, ....
\end{align}
(when $m=0$ we only consider $n= 2, 4,...$).
In the case that $n=|m|+2, |m|+4, ...$, we see that \Autoref{eq:x41}  and \Autoref{eq:x42} gives
\begin{equation}
\label{eq:x43}
\alpha_n^m = \frac{1}{4n}(\beta_{+})_{n-1}^{m+1}+\frac{1}{4n}(\beta_-)_{n-1}^{m-1}-\frac{1}{4(n+2)}(\beta_+)_{n+1}^{m+1}-\frac{1}{4(n+2)}(\beta_-)_{n+1}^{m-1}.
\end{equation}
Now we have from \Autoref{eq:x40}
\begin{align}
\label{eq:x44}
\frac{\partial W}{\partial \nu} &=\sum_{n,m}\frac{1}{2}((\beta_+)_{n}^{m} + (\beta_-)_{n}^{m}) Z_n^m \\
\frac{\partial W}{\partial \mu} &=\sum_{n,m}\frac{1}{2i}((\beta_+)_{n}^{m}-(\beta_-)_{n}^{m}) Z_n^m.
\end{align}
Therefore, from \Autoref{eq:x30} and \Autoref{eq:x31} and the orthogonality/normalization of the $Z_n^m$, see (2.12), we see that the integral expression at the right-hand side of \Autoref{eq:x10} becomes
\begin{align}
\label{eq:x45}
&\frac{1}{\pi}\int_0^1\int_0^{2\pi}\left( \frac{\partial W}{\partial \nu}\frac{\partial U_n^{m*}}{\partial \nu}+ \frac{\partial W}{\partial \mu}\frac{\partial U_n^{m*}}{\partial \mu} \right)\rho d \rho d \theta \nonumber \\
&=-\frac{1}{8\pi}\int_0^1\int_0^{2\pi}\sum_{n',m'}((\beta_+)_{n'}^{m'}+(\beta_-)_{n'}^{m'})Z_{n'}^{m'} \nonumber \\
& \times \left[ Z_{n+1}^{m+1}+Z_{n+1}^{m-1}-Z_{n-1}^{m+1}-Z_{n+1}^{m-1} \right]^* \rho d\rho d \theta \nonumber \\
& =-\frac{1}{8\pi}\int_0^1\int_0^{2\pi}\sum_{n',m'}((\beta_+)_{n'}^{m'}-(\beta_-)_{n'}^{m'})Z_{n'}^{m'} \nonumber \\
& \times \left[Z_{n+1}^{m+1}-Z_{n+1}^{m-1}-Z_{n-1}^{m+1}+Z_{n+1}^{m-1} \right]^*\rho d\rho d \theta \nonumber \\
&=-\frac{1}{8(n+2)}((\beta_+)_{n+1}^{m+1}+(\beta_-)_{n+1}^{m+1}+(\beta_+)_{n+1}^{m-1}+(\beta_-)_{n+1}^{m-1}) \nonumber \\
& + \frac{1}{8n}((\beta_+)_{n-1}^{m+1}+(\beta_-)_{n-1}^{m+1}+(\beta_+)_{n-1}^{m-1}+(\beta_-)_{n-1}^{m-1}) \nonumber\\
&-\frac{1}{8(n+2)}((\beta_+)_{n+1}^{m+1}-(\beta_-)_{n+1}^{m+1}-(\beta_+)_{n+1}^{m-1}+(\beta_-)_{n+1}^{m-1}) \nonumber \\
& + \frac{1}{8n}((\beta_+)_{n-1}^{m+1}-(\beta_-)_{n-1}^{m+1}-(\beta_+)_{n-1}^{m-1}+(\beta_-)_{n-1}^{m-1}),
\end{align}
and this coincides with the right-hand side of \Autoref{eq:x44} when the various cancellations are noted.

In a similar fashion the case $n=|m|>0$ can be handled using \Autoref{eq:x38} and \Autoref{eq:x39}, and this yields that the $\alpha_{|m|}^m$ from \Autoref{eq:x41} and \Autoref{eq:x42} coincide with the right-hand side integral expression in \Autoref{eq:x10}. 
 
\end{document}